\documentclass{article}

\def\giorno{8 April 2002}

\def\*{{\bf ***}}

\def\a{\alpha}
\def\b{\beta}

\def\ga{\gamma}
\def\de{\delta}   

\def\phi{\varphi}

\def\s{\sigma}

\def\om{\omega}

\def\vth{\vartheta}
\def\vphi{\varphi}

\def\G{{\cal G}}

\def\Hb{{\bf H}}
\def\h{{\cal H}}

\def\J{{\cal J}}

\def\K{{\cal K}}
\def\Kb{{\bf K}}
\def\L{{\cal L}}

\def\Ob{{\bf O}}

\def\Q{{\cal Q}}
\def\R{{\bf R}}
\def\S{{\cal S}}
\def\T{{\rm T}}

\def\toro{{\bf T}}
\def\vacca{{\bf V}}

\def\Ga{\Gamma}

\def\La{\Lambda}
\def\Om{\Omega}
\def\Th{\Theta}

\def\pa{\partial}

\def\d{{\rm d}}       
\def\w{\wedge}
\def\xb{{\bf x}}
\def\x{\times}

\def\o+{\oplus}
\def\xd{{\dot x}}

\def\grad{\nabla}     
\def\ss{\subset}
\def\sse{\subseteq}

\def\<{\langle}
\def\>{\rangle}
\def\eor{{$\odot$}}

\def\interno{\hskip 2pt \vbox{\hbox{\vbox to .18
truecm{\vfill\hbox to .25 truecm
{\hfill\hfill}\vfill}\vrule}\hrule}\hskip 2 pt}

\def\({\left(}
\def\){\right)}
\def\[{\left[}
\def\]{\right]}
\def\=#1{\bar #1}
\def\~#1{\widetilde #1}
\def\.#1{\dot #1}
\def\^#1{\widehat #1}
\def\"#1{\ddot #1}

\def\mapright#1{\smash{\mathop{\longrightarrow}\limits^{#1}}}

\def\en#1{\eqno(#1)}
\def\ref#1{\cite{#1}}
\def\Proof{{\bf Proof.}~~}
\def\Remark#1{\medskip \noindent {\bf Remark {#1}}~~}

\begin{document}

\title{\bf Hyperhamiltonian dynamics}

\author{Giuseppe Gaeta\footnote{Supported by ``Fondazione CARIPLO
per la ricerca scientifica''} {~}\footnote{e-mail: gaeta@roma1.infn.it
and gaeta@berlioz.mat.unimi.it} \\
{\it Dipartimento di Fisica, Universit\'a di Roma} \\
{\it Piazzale A. Moro 5, I--00185 Roma (Italy)} \\ and \\
{\it Dipartimento di Matematica, Universit\'a di Milano} \\
{\it via Saldini 50, I--20133 Milano (Italy)} \\ {~~} \\
Paola Morando\footnote{e-mail: morando@polito.it} \\
{\it Dipartimento di Matematica, Politecnico di Torino} \\
{\it Corso Duca degli Abruzzi 24, I--10129 Torino (Italy)} }

\date{{~}}

\maketitle

\noindent {\bf Abstract.} {We introduce an extension of
hamiltonian dynamics, defined on hyperkahler manifolds, which we
call ``hyperhamiltonian dynamics''. We show that this has many of
the attractive features of standard hamiltonian dynamics. We also
discuss the prototypical integrable hyperhamiltonian systems,
i.e. quaternionic oscillators.}

\vfill


\bigskip

{~ \hfill {\it Revised version -- \giorno -- to appear in J. Phys. A}}

\eject
\section*{Introduction}

The description provided by Hamiltonian dynamics applies to many
fields of physics; due to its rich geometrical structure, it is
also very convenient and indeed widely used whenever possible.

Hamiltonian formalism is based on symplectic structures; a
special but relevant class of symplectic manifolds is provided
by Kahler manifolds. Actually, in any symplectic manifold $M$
we can give locally (and globally if $M$ is contractible, e.g.
$M = \R^{2n}$) a complex structure associated to the symplectic
one and, by the introduction of a suitable metric,
a Kahler structure.

In relatively recent years, mathematicians on the one hand, and
(theoretical and mathematical) physicists on the other, have become interested in a special kind of Kahler structures, i.e. {\it hyperkahler} ones \cite{At2}.
These are Kahler with
respect to three different complex structure, having such
relations among them so that, roughly speaking, they can be seen
as the complex structures associated to the three independent
imaginary units of the quaternions  field.

A riemannian manifold (necessarily of dimension $4n$) equipped
with a hyperkahler structure is called a {\it hyperkahler
manifold}. These turn out to be very interesting from the
point of view of Geometry \cite{Qconf,Sal,Swa}, and also relevant in the description of (non-abelian) monopoles \cite{Ati,AtH,Hit1}; they also bear a close connection, which we will not discuss here, with twistors \cite{At2,Pen,War} and thus in particular to interesting classes of integrable systems \cite{Dan,LRU}.
It has also been realized that hyperkahler (and quaternionic-kahler \cite{AlM}) manifolds are relevant in supersymmetry and supergravity theories and related to sigma-models, see e.g. \cite{PHYS,Fre,Hit2}. 
Canonical examples of hyperkahler manifolds are quaternion linear spaces ${\bf H}^n \approx \R^{4n}$ and cotangent bundles of (special) Kahler manifolds \cite{Fre,VeKa}; a specially fruitful method of constructing nontrivial hyperkahler manifolds is through a generalization of Marsden-Weinstein momentum map \cite{Hit2}.
For an overview of recent results in quaternionic and hyperkahler geometry (not needed in the present work), the reader is referred to \cite{Qconf}.

\bigskip

It is obvious that a hyperkahler structure can be described in
symplectic terms; we speak then of a {\it hypersymplectic
structure}. It is remarkable, and came much to our surprise, that
it is possible to provide a generalization of Hamilton mechanics
based on such a hypersymplectic structure, which we do here. What
is more relevant is that this {\it hyperhamiltonian dynamics}
retains most of the appealing features of standard hamiltonian
mechanics, as we show in the present note.

Our initial motivation was provided by integrable systems.
We will consider a special class of these, i.e.
{\it quaternionic oscillators} (see sect. 7), which we
expect to be the paradigm of a nontrivial integrable hyperhamiltonian
system.

Limiting to consider systems with compact energy manifolds, a
$2m$-dimen\-sio\-nal hamiltonian  integrable
system can be described by means of suitable real
(action-angle) coordinates
$(I_a , \phi_a )$, or more compactly complex coordinates $z_a =
I_a \exp [i \phi_a ]$ so that its evolution is described by $m$
constant complex rotations, ${\dot z}_a (t) = i \om_a z (t)$.

In slightly different (but equivalent) terms, we can use
coordinates $(I_a,g_a)$ where $g_a \in G = U(1)$; here of course
$g_a = \exp [i \phi_a ]$ so that we are just describing action
angle coordinates in group-theoretical terms, using the
isomorphism $S^1 \simeq {\bf C}_1 \simeq U(1)$ (here ${\bf C}_1$
are the complex numbers of unit norm). In this
language, the evolution is described by $dI_a /dt = 0$,
$d g_a /dt = \gamma_a$, constant
elements of the Lie algebra $u(1)$; indeed, as well known,
the isomorphism $S^1 \simeq {\bf C}_1 \simeq U(1)$ identifies
the Lie algebra of $U(1)$ with imaginary numbers.

As discussed below, a $4n$-dimensional hyperhamiltonian
integrable system can be described by real (spin) coordinates
$(I_a ; g_a )$, where $g_a \in G = SU(2)$.
Its evolution is accordingly described by $dI_a / d t = 0$,
$d g_a / d t = \gamma_a$, constant elements of the Lie algebra
$su(2)$.

As well known, an integrable  hamiltonian system in a
$2m$ dimensional phase space $M$ is associated to a fibration of
$M$ in tori $\toro^m$. In the case of integrable
hyperhamiltonian systems in a $4 k$ dimensional phase space,
we will not have a fibration in tori
$\toro^{2k} = S^1 \times ... \times S^1
\equiv U(1) \times ... \times U(1)$, but rather in manifolds
$\vacca^k = S^3 \times ... \times S^3 \equiv SU(2) \times ...
\times SU(2)$.

\subsection*{Acknowledgements.}

We would like to thank S. Marchiafava, G. Marmo and M. Tarallo for several discussions at various stages of this work, and the referees for useful suggestions and remarks. 

\section{Quaternionic symplectic structures.}

In this section we recall the basic geometric definitions to be
used in the following. These will mainly concern hyperkahler and
quaternionic geometry; for a short introductions to these the reader is referred to \cite{At2,Cal2}, while more details on these are provided e.g. by \cite{AlM,Ati,AtH,Cal1,Dan,Hit1,Sal,Swa}. Symplectic geometry is discussed e.g. in \cite{EMS4,Can,Fom,GuS};
for hamiltonian mechanics see e.g. \cite{AbM,Arn,EMS3}.

\bigskip

We preliminarly recall that if $M$ is a $2m$ dimensional manifold
equipped with a (riemannian) metric $g$, a {\it complex structure} on
$M$ is a  $(1,1)$ type tensor field $Y$ such that $Y^2 = - I$ which is 
covariant constant; a {\it symplectic form} $\om$ on $M$ is a non-degenerate
and closed two-form.
Then $\om (v,w)$ can be written as $g(v,J w)$ for some $(1,1)$-type
antisymmetric tensor field on $M$. If
this $J$ is orthogonal for $g$, we say that $\om$ is compatible
with the metric (or briefly $g$-compatible), or equivalently that
$\omega$ is {\it unimodular}. If this is the case, then
$(1/m!) \om \w ... \w \om = s \Om$, with $s = \pm 1$ and
$\Om$ the volume form on $M$.
We say accordingly that unimodular symplectic forms are
of positive or negative type.

We also recall that $M$ is a {\it Kahler manifold} if it is
equipped with a metric $g$, a complex structure $Y$ and a
symplectic form $\om$, satisfying the {\it Kahler relation} $\om
(v,w) := g (v,Yw)$, or equivalently  $g (v,w) = \om (Yv,w)$.

In local coordinates, if $Y^i_{~j} (x)$ describes the complex
structure, the associated symplectic form is given by $\om = (1/2) K_{ij}
(x) \d x^i \w \d x^j$ with $K_{ij} (x) = g_{im} (x) Y^m_{~j} (x)$.

\bigskip

We can pass to consider hyperkahler manifolds. Now -- and
always in the following -- $M$ will be a smooth $4n$ dimensional
real manifold endowed with a riemannian metric $g$, and $\epsilon$ will denote the completely antisymmetric (Levi-Civita) symbol.

\medskip\noindent
{\bf Definition 1.} {\it A {\rm hypercomplex structure} on $M$ is
an ordered triple ${\bf Y} = (Y_1,Y_2,Y_3)$ of complex structures
on $M$ satisfying $Y_\a Y_\b = \epsilon_{\a \b \ga} Y_\ga -
\delta_{\a \b} I$. If the $Y_\a$ are orthogonal complex
structures on $(M,g)$, we say that ${\bf Y}$ is orthogonal.}

\Remark{1.} The $Y_\a$ making up a hypercomplex structure satisfy
the quaternionic relations. In Lie algebraic terms, the $\~Y_\a = (1/2) Y_\a$, which satisfy $[\~Y_\a , \~Y_\b ] = \epsilon_{\a \b \ga} \~Y_\ga$, realize the $su(2)$ algebra. $\odot$

\medskip\noindent
{\bf Definition 2.} {\it A {\rm hyperkahler structure} on $M$ is
a quadruple $(g,Y_1,Y_2, Y_3)$ where: $g$ is a metric on $M$; the
$Y_\a$ are an orthogonal hypercomplex structure $(M,g)$; and the
two-forms $\om_\a$ defined by the complex structures $Y_\a$ via
the Kahler relation are closed and nondegenerate on $M$.}
\medskip

Notice that the forms $\om_\a$ are therefore (independent)
symplectic forms on $M$. As dealing with differential forms is
equivalent to -- but rather more convenient in practice -- than
dealing with $(1,1)$ tensor fields, we will generally find more
convenient to focus on these.

\medskip\noindent
{\bf Definition 3.} {\it A {\rm hypersymplectic structure}
$\Ob = \{ \om_1 , \om_2 , \om_3 \}$ on the riemannian manifold
$(M,g)$ is an ordered triple of $g$-compatible symplectic
structures on $M$, such that the complex structures $Y_\a$
defined by the $\om_\a$ via the Kahler relation are a
hypercomplex structure on $M$.}
\medskip

If $\Ob$ is a hypersymplectic structure on $(M,g)$, the linear
span (with real coefficients) of the $\om_\a$ is the real linear
space $\Q :=  \{ \sum_\a c_\a \om_\a  \} \ss \Lambda^2 (M)$. This
is called the {\it quaternionic symplectic structure } generated
by the hypersymplectic structure $\Ob$, and $\Ob$ is an
{\it admissible basis} for $\Q$
\cite{AlM}. The unit sphere in $\Q$ (with the natural metric, see
below) will be denoted as $\S$ ($\S \approx S^2$).

\Remark{2.} The $\S$ defined above is related to the {\it
twistor space} on $M$ \cite{Ati,AtH,At2,Hit1,Pen,War}. \eor
\medskip

The natural scalar product in $\Q$ (seen as a linear space)
between $q_1 = a_\a \om_\a$ and $q_2 = b_\a \om_\a$ is $(q_1,q_2)
:= a_\a b_\a$. If we choose  a local coordinates system, and we associate
to $q_i$ the matrices $Q_i$, this
coincides with the natural scalar product in the linear space
${\bf Q}$ generated by the complex stuctures $\{ Y_\a \}$, i.e. $(Q_1,Q_2)
:= (4n)^{-1} {\rm Tr} (Q_1^\dagger Q_2) $.

\medskip\noindent
{\bf Lemma 1.} {\it Any nonzero $\om \in \Q$ is a symplectic
structure on $M$. If $\om \in \Q$, then $\om$ is unimodular (and
thus defines a Kahler structure in $M$) if and only if $\om \in
\S$.}

\medskip\noindent
\Proof The first part is trivial. As for the second, if $\om =
c_\a \om_\a \in \Q$, then $\om$ yields the complex structure $Y =
c_\a Y_\a$, where the $\{ Y_\a \}$ are the hyperkahler structure
(so that, in particular, $\{ Y_\a , Y_\b \} = - 2 \delta_{\a \b}$
). We have therefore $Y^2 = - Y^T Y = - (\sum_\a c_\a^2 ) I$, and
hence the statement. $\ \triangle$
\medskip

The three symplectic structures $\om_\a$ can be seen as
associated to the imaginary units of the quaternions; it is thus
natural that if we operate a (pure imaginary) rotation in the
quaternions, we obtain three different symplectic structures
which still generate the same quaternionic structure. In other
words, we can change the basis in $\Q$ preserving the
quaternionic relations, i.e. passing to a different admissible
basis. Notice that in this case the sphere $\S \ss \Q$ is
invariant.

\medskip\noindent
{\bf Definition 4.} {\it Two hypersymplectic structures $(\Ob,g)$
and $(\^\Ob , g)$ on $M$, spanning the same quaternionic
symplectic structure are said to be {\rm equivalent}.}

\Remark{3.} More generally, consider a map $\Phi : M \to M$ and
let $\Phi^*$ be its pullback; if we consider local coordinate 
$\{x^i\}$ on $M$, we can write $\om = (1/2) K_{ij} (x) \d x^i
\w \d x^j$; then $\Phi^* (\om ) = (1/2) (A^+ K A)_{ij} (x) \d x^i
\w \d x^j $, where $A = (D \Phi)$ is the jacobian of $\Phi$. Thus
if $A (x) \in O (4n , \R)$, then $\Phi^*$ is a morphism of
hyperkahler structures of $(M,g)$; if $A^+ {\bf Q} A = {\bf Q}$, then
$\Phi^*$ maps $\Q$ into itself (and necessarily preserves $\S$),
i.e. maps $\Ob$ to an equivalent hypersymplectic structure. \eor

\section{Equations of motion.}

In this section we define a class of equations of motion in a
hyperkahler manifold; these are associated to the hyperkahler
structure and define a Liouville dynamics on the manifold.

Let us consider a hyperkahler manifold $(M,g,Y_1,Y_2,Y_3)$ of
real dimension $4n$. This can be equivalently seen as a
hypersymplectic manifold
$(M,g,\om_\a)$ with $\om_\a$ the symplectic forms associated
to $Y_\a$ via $g$; in the following we will refer to the
hyperkahler structure even when we will focus on the
symplectic aspect. The symbol $s$ will have value $\pm 1$,
depending if we are considering positive or negative type symplectic forms
$\om_\a$. We define, for ease of notation, $\zeta_\a = \om_\a \w
... \w \om_\a$ (with $2n-1$ factors).

Any triple of smooth functions $\h^\a : M \to \R$ ($\a=1,2,3$),
defines a vector field $X: M \to \T M$ by the
equations of motion
$$ X \interno \, \Omega \
= \frac {1}{(2n-1)!}\sum_{\a=1}^3 \, \d \h^\a \w \zeta_\a \ .
\eqno(1)
$$
We call this the {\it hyperhamiltonian vector field} on $(M,g,\Ob )$
associated to the triple $\h^\a$.

We stress that the vector field $X$ is uniquely defined by this. Also
note that, for any $\a$, one gets $ \om_\a \w ... \w \om_\a  =  [(2n)!] \,
s \, \Om$ (the $\om_\a$ involved in the wedge product are  $2n$).
 Using this relation, the equations of motion can also
be rewritten as
$$ X \interno \sum_{\a=1}^3 \, \om_\a \w \zeta_\a \ = \ (6 s  n)
\ \sum_{\a=1}^3 \, \d \h^\a \w \zeta_\a \ . $$

\medskip\noindent
{\bf Lemma 2.} {\it Equation (1) defines a Liouville vector
field on $M$, i.e. $\L_X ( \Om ) = 0$.}

\medskip\noindent
{\bf Proof.} By the general definition of Lie
derivative, $\L_X (\Om ) = \d (X \interno \Omega ) + X \interno
(\d \Om )$, and obviously $\d \Om = 0$. If $X$ satisfies (1) we
have therefore $\L_X (\Om ) = \sum_\a \d ( \d \h^\a \w \zeta_\a
)$, which is zero since the forms $\zeta_\a$ are closed.
$ \ \triangle$
\medskip

\medskip\noindent
{\bf Lemma 3.} {\it The vector field $X$ defined by (1) can be rewritten as
$X = X_1 + X_2 + X_3$, where $X_\a$ satisfies $X_\a \interno \om_\a = \d \h^\a$
for $\a = 1,2,3$.}

\medskip\noindent
{\bf Proof.} It is immediate to check that $X = \sum_\a X_\a$ with $X_\a$
defined as above yields a solution to (1). As $X$ is uniquely defined by (1),
this proves the statement.
$ \ \triangle$
\medskip

\Remark{4.} It follows from these that if $\om_\a = (1/2) K^{(\a)}_{ij}
\d x^i \w \d x^j$, then the hyperkahler
vector field (1) can be written as $X = f^i \pa_i$ where
$ f^i = \sum_{\a} \, K_\a^{ij} \grad_j \h^\a $ \eor
\medskip

\Remark{5.} If we have a hyperkahler manifold $(M,g)$
and an hamiltonian vector
field $X$ (with respect to a symplectic structure $\om$ being part of a hypersymplectic structure $\bf O$), this can obviously be seen as a hyperhamiltonian vector field just by choosing two of the $\h^\a$ to be constant.
One could wonder if all the hyperhamiltonian vector fields on $M$ can be hamiltonian by a suitable choice of a symplectic structure; this is not the case even in the simplest setting ($M = \R^4$), as we show by
explicit example in lemma 5 (see section 6). \eor
\medskip

The hyperhamiltonian vector field in $M$ induces a vector field (which we also call hyperhamiltonian) in extended phase
space, i.e. in $M \x \R$, where the $\R$ space has coordinate $t$
(and represents the time).

If we introduce local coordinates $\{x^1, \ldots x^{4n},t\}$ in $M \times {\bf R}$,
the dynamics in $M \x \R$ will be described by a vector field $
Z  =  z^0 (x,t) \pa_t  + z^i (x,t) \pa_i$ (here and in the
following we write $\pa_i$ for $\pa / \pa x^i$, $\pa_t = \pa /
\pa t$). The equations of motion given above are equivalent to
defining the vector field $Z$ to be
$ Z  =  \pa_t  +  X $, where obviously $X$ is
defined by (1).

Due to the closeness of $\om_\a$, we can locally find
one-forms $\s_\a$ such that $\om_\a = \d \s_\a$, and locally 
define forms $\vphi,\vth$ in $\Lambda^{(4n-1)} (M \x \R)$ given by
$$ \vphi \ = \ \sum_{\a=1}^3 \, \s_\a \w \zeta_\a  \quad , \quad
\vth \ = \ \vphi \ + (6 s  n)\ \sum_{\a=1}^3 \, \h^\a \ \zeta_\a
\w \d t \ . \eqno(2) $$ Note, for later use, that the $(4n)$-form
$\d \vth$ is nonsingular, and that $\d \vphi$ is proportional to
the volume form $\Om$. 

When $\om_\a$ is exact (in particular if $M$ has vanishing second cohomology group, e.g. for $M = \R^{4n}$), the $\s_\a$ and related forms are globally defined. In order to avoid repeating too frequently that the considerations to be presented are local, we will assume from now on that the $\om_\a$ are exact.

\medskip\noindent
{\bf Theorem 1.} {\it Let $M$ be a hyperkahler manifold, and let
$\h^\a : M \to \R$ ($\a = 1,2,3$) be assigned smooth functions;
let $\vth$ be the form defined by (2). Then the equations of
motion (1) are equivalent to $$ Z \interno \d \vth \ = \ 0 \ \ ,
\ \ Z \interno \d t \ = \ 1 \eqno(3) $$ where $Z$ is a vector
field on $M \x \R$.}

\medskip\noindent
\Proof The equation $Z \interno \d t = 1$ means that we can write
$Z$ in the form $Z = \pa_t + Y$; the other equation $Z \interno
\d \vth  =  0$ yields then, with simple algebra and separating
forms with and without a $\d t$ factor, two equations:
$$ Y \interno \sum_{\a=1}^3 \, \om_\a \w \zeta_\a \, =  \, (6 s  n) \,
 \sum_{\a=1}^3 \d \h^\a \w \zeta_\a \ \ {\rm and} \ \
 \, Y \interno \sum_{\a=1}^3 \d \h^\a \w \zeta_\a \, = \, 0 \ .
$$ The second of these is a trivial consequence of the first one;
but the first is just (1). Thus, as (1) uniquely determines $X$,
we have $Y \equiv X$. $\ \triangle$

\section{Conservation laws and Poisson-like brackets}

For the class of systems defined above we have a natural
conserved $(4n-1)$-form $\Theta$, canonically associated to the
triple $\{ \h^1 , \h^2 , \h^3 \}$ and defined as
$$ \Th \ :=  \, \sum_{\a=1}^3 \d \h^\a \w \zeta_\a \ . \eqno(4) $$

\medskip\noindent
{\bf Theorem 2.} {\it Let $(M,g,\om_\a)$ be a hyperkahler
manifold, $\{ \h^\a \}$ be any triple of functions $\h^\a : M \to
\R$, $X$ be the hyperhamiltonian flow defined by (1), and
$\Theta$ defined by (4). Then $\L_X (\Th)=0$.}

\medskip\noindent
\Proof The form $\Th$ is closed, hence $ \L_X (\Th ) = \d (X
\interno \Th)$; the explicit expression of $\Th$ and the
equations of motion (1) give
$$ (X \interno \Th) \ = \ (2n-1)! \ \[X \interno
\( X \interno \Om \) \] \ , $$
which is identically zero since we are contracting twice an
alternating form with the same vector. $\ \triangle$
\medskip

Notice that $(4n-1)$ forms $\chi$ on $M$ are canonically
associated to vector fields $Y$ on $M$ via $Y \interno \Omega =
\chi$; we write $Y = F (\chi )=Y_\chi $. There is a natural
operation $\{ . , . \} : \Lambda^{(4n-1)} (M) \times
\Lambda^{(4n-1)} (M) \to \Lambda^{(4n-1)} (M) $, defined as
follows. Given forms $\chi , \Psi \in \Lambda^{(4n-1)} (M) $, we
consider the associated vector fields $Y_\chi , Y_\Psi$; take the
commutator $ Y_\Ga := [ Y_\chi , Y_\Psi ]$. This defines an
associated form $\Ga \in \Lambda^{(4n-1)} (M)$, and we define $\{
\chi , \Psi \}$ to be just $\Ga$. In other words,
$$ \{ \chi , \Psi \} \ := \ F^{-1} \( \, [ F (\chi ) , F(\Psi ) ]
\, \) \ . \eqno(5)$$

We stress that with this notation, $\Theta = F (X)$ with $X$ the
hyperhamiltonian vector field. Note also that if we have two
conserved $(4n-1)$-forms $\Theta_i$, we can generate another
(possibly not independent from these, or zero) conserved form $\{
\Theta_1 , \Theta_2 \}$. In this respect, $\{ . , . \}$
is reminiscent of the Poisson brackets of standard Hamiltonian
mechanics; however we have to remark that the situation differs substantially from the standard Poisson brackets, because to define our brackets we don't use the hyperkahler structure of the manifold, but only the isomorphism between vector fields and $(4n-1)$-forms  induced by the volume form on $M$.

\section{Variational formulation}

In this section we will formulate a local variational principle related to the hyperhamiltonian equations of motion introduced in section 2. 

In order to state in a geometrical framework our principle, we need to consider a local fiber bundle structure on the hyperkahler manifold $M$. This local fibration allows to describe a particular class of variations (``vertical'' with respect to the fibration, as we precise in a moment) that generalize isochronous variations considered in the variational principle for standard hamiltonian mechanics.
\medskip

Let us consider a hyperkahler manifold $(M,g,\om_\a)$ of real
dimension $4n$, and a triple $\{ \h^\a \}$ of hamiltonian
functions; we will consider the extended phase space $M \x \R$,
which we see as a trivial fiber bundle $t: M \x \R \to \R$.

In order to properly set the local variational problem in a chart  $M_i$ of $M$, we will 
need to consider a double fibration 
$$ M_i \x \R \ \mapright{\pi_i} \ B_i \ \mapright{\tau_i} \ \R $$ 
where the base manifold $B_i$ of the fiber
bundle $\pi_i : M_i \x \R \to B_i$ is a manifold of dimension $(4n-1)$, fibered itself 
over $\R$ with projection $\tau_i$. We also require, obviously, that $\tau_i \circ \pi_i = t$ 
on $M_i \x \R$.

For ease of notation, we will from now on just write $M$ for $M_i$ and $B$ for $B_i$, i.e. use 
a ``global'' notation. We stress that the double fibration considered here is not a general 
global construction associated to the geometrical structure of an hyperkahler manifold; 
anyway, this double fibration can be considered locally in $M_i$ for a generic hyperkahler 
manifold $M$, and the choice of the local base manifold $B_i$ is widely arbitrary. Note that 
in the simple but relevant case $M = \R^{4n}$ the double fibration exists globally.

We denote the sets of sections of the bundles introduced above, respectively, by 
$\Gamma (\pi )$ and $\Gamma (\tau)$; and similarly for $\Gamma (t)$. We denote by 
${\cal V} (\pi) $ the set of vertical vector fields for the fibration $\pi : M \x \R \to B$.

For $V \in {\cal V} (\pi)$, we denote by $\psi_s : M \x \R \to M
\x \R$ the flow generated by $V$. We want to consider variations
of sections\footnote{Should the reader be misled by our ``global'' notation, we note 
these are actually local sections $\Phi_i \in \Ga (\pi_i)$, i.e. $\Phi_i : M_i \to M_i \x \R$.} 
$\Phi \in \Ga (\pi)$ under the action of $V \in {\cal V} (\pi)$ \cite{Her,Sau}.

\medskip\noindent
{\bf Definition 5.} {\it Let $\Phi \in \Gamma (\pi)$. The {\rm
variation} of $\Phi$ under the vertical vector field $V$ is the
section $ \~\psi_s (\Phi)  :=  \psi_s \circ \Phi \ \in \Gamma
(\pi)$.}
\medskip

\Remark{6.} The nature of the double fibration $M \x \R
\mapright{\pi}  B  \mapright{\tau}  \R$, where $\tau \circ \pi =
t$, ensures that vertical vector fields $V \in {\cal V}$ cannot
have components along $\pa_t$; that is, we are actually
considering isochronous variations. We also recall that, in order 
to consider variation of the section $\Phi$, we don't need a vertical 
vector field $V$ defined on all $M \times {\bf R}$, but just a vertical 
vector field defined along $\Phi$. $\odot$
\medskip

The main object to be considered is the $(4n -1)$-form $\vth$ on
$M \x \R$, defined by (2). We recall that $ \vth :=
\sum_{\a=1}^3  \[ \s_\a \w \zeta_\a \, + \,(6 n s)  \h^\a \,
\zeta_a \w \d t \]$, where $ \d \s_\a = \om_\a$.

Let us consider a compact $(4n-1)$ dimensional submanifold 
with boundary $C \sse B$. 
We define a functional $I  : \Gamma (\pi ) \to \R$ given by
$$ I  (\Phi ) \ := \ \int_C \, \Phi^* (\vth ) \eqno(6) $$
where $\Phi^* (\vth)$ denotes, as customary, the pullback of $\vth$ by $\Phi$.

\medskip

In the following we will consider only vertical vector fields $V \in {\cal V} (\pi)$ 
such that $V$ vanish on $\pi^{-1} (\pa C)$, where $\pa C$ is the boundary of $C$. 
This is just the familiar condition of zero variation on the boundary of the integration region. We denote these as ${\cal V}_C (\pi)$.

\medskip\noindent
{\bf Definition 6.} {\it A section $\Phi \in \Gamma (\pi)$ is {\rm
extremal for $I $} if and only if
$$ {\d ~ \over \d s} \ \[ \int_C \, \( \~\psi_s (\Phi ) \)^*
(\vth) \]_{s=0} \ = \ 0 \eqno(7) $$ 
whenever $V \in {\cal V}_C (\pi)$. In this case we write 
$(\de I ) (\Phi)=0$.}

\medskip\noindent
{\bf Theorem 3a.} {\it A section $\Phi \in \Gamma (\pi)$ is
extremal for $I $ defined by (6)  if and only if $ \Phi^* ( V \interno \d \vth )  =  0$ for all  $V \in {\cal V}_C (\pi)$.}

\medskip\noindent
\Proof This is a standard theorem of variational analysis, see
e.g.  chapter XII of \cite{Her}. $\ \triangle$
\medskip

\Remark{7.} Note that ${\cal V} (\pi ) $ is two dimensional as a
module over the algebra of smooth functions $F : M \x \R \to \R$.
With $V_1,V_2$ a pair of generators for ${\cal V} (\pi)$, 
the condition $ \Phi^* ( V \interno \d \vth ) = 0 $ $\forall V \in {\cal V}_C (\pi)$ can be written as $\Phi^* \( V_1 \, \interno \, \d \vth \)  = 0 = \Phi^* \( V_2 \, \interno \, \d \vth \)$. This is independent of $C$. $\odot$
\medskip

Sections $\Phi$ which are extremal for $I $ are related to the hyperhamiltonian vector field $Z$ in that $Z$ is the characteristic vector field for $\Phi$, as discussed below.  
\medskip

The relation between $I $ and $Z$ is better understood in the language of ideals of differential forms \cite{Car,God} (some basic definitions used here are recalled in the appendix), which we will call just ideals for short. With this language, and recalling remark 7, theorem  3a above can be restated as follows: 

\medskip\noindent
{\bf Theorem 3b.} {\it Let $V_1,V_2$ generate ${\cal V} (\pi)$. 
A section $\Phi \in \Gamma (\pi)$ is extremal for $I $ defined by (6) if and only if $\Phi$ is an integral manifold of the ideal $\J$ generated by $V_1 \interno \d \vth$ and $V_2 \interno \d \vth$.}
\medskip

In view of this fact, we will say that $\J$ is the ideal associated to 
the variational principle $\de I  = 0$.

We can now discuss the relation between the vector field $Z$ introduced in section 2 and the variational principle based on $I $.
We will first establish a simple lemma and an immediate corollary thereof.

\medskip\noindent
{\bf Lemma 4.} {\it Let $\a$ be a nonzero $N$-form in the $(N+1)$-dimensional manifold $M$. 
Let $X,V_1,V_2$ be three independent and nonzero vector fields on $M$. 
Then $V_1 \interno (X \interno \a) = 0  = V_2 \interno (X \interno \a) = 0$ implies (and is thus equivalent to) $X \interno \a = 0$. 
Moreover, the space of vector fields $Y$ satisfying $Y \interno \a = 0$ is a one dimensional module over $\Lambda^0 (M)$.}

\medskip\noindent
{\bf Proof.} Choose local coordinates $\{x^0 , x^1 , ... , x^N \}$ in $M$; we can always 
take $X = \pa_0$, $V_1 = \pa_1 $, $V_2 = \pa_2$. 
We write $\Om = \d x^0 \w ... \w \d x^N$; then, in full generality,  
$ \a = \sum_{k=0}^N c_k (\pa_k \interno \Om )$.
Now $\pa_1 \interno (\pa_0 \interno \a) = 0$ implies $c_j = 0$ for $j \not= 0,1$; 
and $\pa_2 \interno (\pa_0 \interno \a) = 0$ implies $c_j = 0 $ for $j \not= 0,2$. 
Imposing both equations yields $\a = c_0 (\pa_0 \interno \Om) \equiv c_0 (X \interno \Om)$. 
This satisfies, of course, $X \interno \a = 0$; conversely $Y \interno \a = 0$ implies 
$Y = f X$. $\triangle$

\medskip\noindent
{\bf Corollary.} {\it Let $\a,V_1,V_2$ be as above. Then the ideal $\J$ generated by 
$\{ \Psi_1 = (V_1 \interno \a), \Psi_2 = (V_2 \interno \a) \}$ is nonsingular and admits a one-dimensional characteristic distribution $D(\J )$; this is given by vector fields satisfying $X \interno \a = 0$.}

\medskip\noindent
{\bf Proof.} As $\Psi_1, \Psi_2$ are both $N-1$ forms, $(X \interno \Psi_j) \in \J$ is 
equivalent to $X \interno \Psi_j = 0$. 
Thus the corollary is merely a restatement of lemma 4; notice this implies that the space of vector fields satisfying $X \interno \a = 0$ has constant dimension, i.e. $\J$ is nonsingular. $\triangle$

\medskip\noindent
{\bf Theorem 4.} {\it Let $(M,g,\om_\a)$ be a hyperkahler manifold
of real dimension $4n$; let $\{ \h^\a : M \to \R \}$  be three
smooth functions. Let $\vth$ be the $(4n-1)$-form de\-fi\-ned by (2),
and let $\J$ be the nonsingular ideal associated to the variational principle defined by $I $.
Then the characteristic distribution $D (\J)$ for $\J$ is one-dimensional and is generated by the hyperhamiltonian vector field  $Z$ defined by (3).}

\medskip\noindent
\Proof Specialize lemma 4 and its corollary to the case 
$\a = \d \vth$, and use theorem 1. $\triangle$
\medskip

\Remark{8.} It follows from this that the vector field $Z$ is everywhere tangent to integral manifolds of $\J$, i.e. to extremal sections for $I $. Moreover, by proposition A1 (see the appendix), it also shows that the $(4n-1)$-dimensional extremal sections $\Phi$ 
for $I $ can be described by assigning their value on a suitable $(4n-2)$-dimensional manifold and pulling them along integral curves of $Z$. $\odot$

\section{Integral invariants.}

The Poincar\'e invariants (Poincar\'e form and Poincar\'e-Cartan
integral invariant) play a central role in the canonical
structure of Hamiltonian mechanics. In hyperhamiltonian
mechanics, we have objects enjoying the same properties; these
turn out to be, respectively, the forms $\vth$ and $\vphi$
introduced above, see (2).

We consider as usual a differentiable manifold $M$ of dimension
$4n$ equipped with a hypersymplectic structure $\{ \om_\a \}$,
and the extended phase space $M \times \R$. 

\medskip\noindent
{\bf Theorem 5.} {\it Let $\ga_0$ be a closed and oriented
$(4n-1)$ submanifold of the extended phase space $M \x \R$; 
let $\ga_t$ be the manifold obtained by transporting $\ga$ along the flow of the hyperhamiltonian vector field $Z$ defined in (3).
Then
$$ {\d \over \d t} \ \int_{\ga_t} \ \vth \ = \ 0 \ . \eqno(8) $$}

\medskip\noindent
\Proof Let $Z_t$ denote the flow of $Z$. We have
$$ {\d \over \d t} \, \int_{\ga_t} \, \vth \ = \ {\d \over \d t} \, \int_{\ga_0} \, Z_t^* \vth  \, =  \, \int_{\ga_0} \, {\d \over \d t} ( Z_t^* \vth ) \, =  \, \int_{\ga_0} \, \ Z_t^* [ d(Z \interno \vth) + Z \interno \d \vth ] \, = \, 0 \ , \eqno(9) $$
where the last integral vanishes because $Z  \interno \d \vth \ = \ 0$ and  (by Stokes' theorem) because $\ga_0$ is closed.
 $ \ \triangle$ 
\medskip

We consider the special case of the construction considered above
(leading to the Poincar\'e-Cartan invariant) in which the manifold
$\ga_0$ lie on a hyperplane at $t$ constant. 

This gives the Poincar\'e relative invariant, that should also be reinterpreted as the conservation of the volume form under the hyperhamiltonian flow.

\medskip\noindent
{\bf Theorem 6.}  {\it Let $\ga_0$ be a closed and oriented
$(4n-1)$ submanifold of the extended phase space $M \x \R$ lying
in the fiber over $t_0$ of the fibration $t:M \x \R \to \R$. let $\ga_t$ be the manifold obtained by transporting $\ga$ along the flow of the hyperhamiltonian vector field $Z$ defined in (3).
Then
$$ {\d \over \d t} \ \int_{\ga_t} \ \vphi \ = \ 0 \ . \eqno(10) $$}

\medskip\noindent
\Proof  In this case $\d t = 0$ on $\ga_t$ and the integration of
$\vth$ on $\ga_t$ reduces to the integration of $\vphi$ on the
same manifold $\ga_t$. Therefore (8) means that the integral of
$\vphi$ over $\ga_t$ is constant, i.e. (10). $\ \triangle$
\medskip

\section{Hypersymplectic structures in $\R^4$.}

\subsection{Standard structures}

After developing the general theory in abstract terms, it will be
useful to consider the simplest nontrivial example of
hypersymplectic manifold. This is provided by $M = \R^4$ with
standard euclidean metric $g_{ij} (x) = \delta_{ij}$.

Despite the fact this is just a (simple) exercise of linear algebra, we will explicitly write the hyperhamiltonian equations, and this for three reasons: (a) this is the simplest case in which our construction applies, and having a fully explicit example can only help our understanding; (b) the explicit formulation of the equation of motion in the standard case will be useful in Section 7 when we discuss quaternionic oscillator; (c) discussion of this simple case will clarify some points which were not fully discussed above, referring instead to this explicit example. 

The latter were: first, the reason why we left the possibility that the orientation of the metric and the orientation of the symplectic structure disagree; and second, we use the standard structure to give an explicit example of a hyperhamiltonian vector field that is not hamiltonian, whatever symplectic structure we define on $M$, showing that the dynamic that we propose is a real extension of Hamiltonian dynamics.

\bigskip

We use cartesian coordinates $x^i$ in ${\bf R}^4$, and the volume form will be $\Om = \d x^1 \w \d x^2 \w \d x^3 \w \d x^4$. The space $\La^2 (\R^4)$ is six dimensional, and is spanned by
$$ \begin{array}{rlcrl}
\mu_1  =& \ \d x^1 \w \d x^2 + \d x^3 \w \d x^4 & \ , \ &
\eta_1 =& \ \d x^1 \w \d x^3 + \d x^2 \w \d x^4 \ , \\
\mu_2  =& \ \d x^1 \w \d x^4 + \d x^2 \w \d x^3 & \ , \ &
\eta_2 =& \ \d x^4 \w \d x^1 + \d x^2 \w \d x^3 \ , \\
\mu_3  =& \ \d x^1 \w \d x^3 + \d x^4 \w \d x^2 & \ , \ & \eta_3
=& \ \d x^2 \w \d x^1 + \d x^3 \w \d x^4 \ . \end{array}$$ Note
that the $\mu$ span the space $\Lambda_+^2 (M)$ of self-dual
forms, the $\eta$ span the space $\Lambda_-^2 (M)$ of
anti-self-dual forms.

Note also that the $\mu_\a \w \mu_\a = \Om$, $\eta_\a \w \eta_a =
- \Om$. We will refer to these as standard hypersymplectic
structures of positive and negative type respectively. We also
denote as $\Q_\pm$ the quaternionic structures spanned by these,
and $\S_\pm$ their unit spheres. Obviously we have $\Lambda_\pm^2
(M) = \Q_\pm$.

We write two-forms on $\R^4$ as $\om = (1/2) (J)_{im} \d x^i \w
\d x^m$, with $J$ an antisymmetric tensor. We write the tensors
corresponding to the $\mu_\a$ as $K_\a$, those corresponding to
$\eta_\a$ as $H_\a$ (and their triples as $\Kb$ and $\Hb$).

Explicit expressions of these are as follows:
$$ K_1 = \pmatrix{0&1&0&0\cr
-1&0&0&0\cr 0&0&0&1\cr 0&0&-1&0\cr} ~,~ K_2 = \pmatrix{0&0&0&1\cr
0&0&1&0\cr 0&-1&0&0\cr -1&0&0&0\cr} ~,~ K_3 = \pmatrix{0&0&1&0\cr
0&0&0&-1\cr -1&0&0&0\cr 0&1&0&0\cr} \ . $$

$$
H_1 = \pmatrix{0&0&1&0\cr 0&0&0&1\cr -1&0&0&0\cr 0&-1&0&0\cr} ~,~
H_2 = \pmatrix{0&0&0&-1\cr 0&0&1&0\cr 0&-1&0&0\cr 1&0&0&0\cr} ~,~
H_3 = \pmatrix{0&-1&0&0\cr 1&0&0&0\cr 0&0&0&1\cr 0&0&-1&0\cr} ~.
 $$

The complex structures are $Y_\a = g^{-1} J_\a$ and in the present
case of euclidean metric we also write them as $K_\a$ and $H_\a$
(with a raised index).

\Remark{9.} The $\Kb$ and $\Hb$ span $su(2)$ algebras, which
we denote as $su(2)_\pm$; they correspond to the left and right
spinor algebras (note indeed that reversing the orientation of
space exchanges the $\Kb$ and $\Hb$). $\odot$

It is immediate to check that $[ K_\a , H_\b ] = 0$ for all
$\a,\b$;  actually, if we look for the centralizer of $SU(2)_\pm$
in $GL(4,\R)$, this is just $SU(2)_\mp$. This corresponds to the
well known relation $so(4) \simeq su(2)_+ \oplus su(2)_-$, or
equivalently to $\Lambda^2 (M) = \Lambda^2_+ (M) \oplus
\Lambda^2_- (M) \equiv \Q_+ \oplus \Q_-$.

In the case of the standard positive-type hypersymplectic structure
in $M=(\R^4 , \delta )$,  the equations of motion will be simply
$$ {\dot x}^i \ = \ \ \sum_{\a = 1}^3 \ (K_\a)^{ij}
\, {\pa \h^\a \over \pa x^j } \ . \eqno(11) $$
For the negative-type hypersymplectic structure the $K_\a$ are replaced by the $H_\a$.

\medskip
{\bf Lemma 5.} {\it There are equations of the form (11)
which are not hamiltonian, whatever symplectic structure we define
in $M$.}

\medskip
{\bf Proof.}
To prove this, we use the following result from sect.3 of
\cite{Mar}: {\it Given a linear vector field $X = A^i_{~j} x^j
\pa_i$, if ${\rm Tr} (A^{2k +1}) \not=0$ for some $k \in {\bf
N}$, this is not hamiltonian with respect to any symplectic
structure}. Note that vanishing of ${\rm Tr} (A)$ corresponds to
the condition of zero divergence, which is also satisfied by
hyperhamiltonian flows.

Thus we only have to exhibit an example where $\h^\a = (1/2)
D^\a_{ij} x^i x^j$ (with $D^\a$ symmetric matrices) and $A :=
\sum_\a \, K_\a D^\a$ satisfies ${\rm Tr} (A^3) \not= 0$. This is
obtained e.g. if $\h^1 = (1/2) [(x^1)^2 - (x^2)^2 + (x^3)^2 -
(x^4)^2 + 2 (x^1 x^4 - x^2 x^3)]$, $\h^2 = (1/2) |x|^2$, and
$\h^3 = 0$. $\ \triangle$
\medskip

We will now consider the standard
hypersymplectic structures defined above, and
derive explicit expressions for the associated hyperhamiltonian
dynamics. We consider the positive-type hypersymplectic structure.

It is immediate to check that, for any $\a = 1,2,3$,
$$ \om_\a \w \om_\a \ = \ 2 \ \Om \ = \
2 \ \d x^1 \w \d x^2 \w \d x^3 \w \d x^4. $$
Obviously, with $X = f^i \pa_i$, we
have
$$ \begin{array}{rl}
X \interno \Om \ = \ & f^1 \, \d x^2 \w \d x^3 \w \d x^4 \ - \
f^2 \, \d x^1 \w \d x^3 \w \d x^4 \ + \\ + & \ f^3 \, \d x^1 \w \d x^2
\w \d x^4 \ - \ f^4 \, \d x^1 \w \d x^2 \w \d x^3 \ . \end{array}$$
The computation of $\d \h^\a \w \om_\a$ is also immediate, and we get
$$ \begin{array}{rl}
\sum_{\a=1}^3 \d \h^\a \w \om_\a \ = & \ \ (\pa_3 \h^1 + \pa_1 \h^2 - \pa_2 \h^3 ) \, \d x^1 \w \d x^2 \w \d x^3 \ + \\
 & + \ (\pa_4 \h^1 - \pa_2 \h^2 - \pa_1 \h^3 ) \, \d x^1 \w \d
x^2 \w \d x^4 \ + \\
 & + \ (\pa_1 \h^1 - \pa_3 \h^2 + \pa_4 \h^3 ) \, \d
x^1 \w \d x^3 \w \d x^4 \ + \\
 & + \ (\pa_2 \h^1 + \pa_4 \h^2 + \pa_3
\h^3 ) \, \d x^2 \w \d x^3 \w \d x^4 \ . \end{array} $$
The equations of motion are immediately obtained by comparing this
and the previous expression; these are
$$ \cases{ {\dot x}^1 \ = \  (\pa \h^1 / \pa x^2)  + (\pa \h^2 / \pa x^4) +
(\pa \h^3 / \pa x^3)  & \cr
{\dot x}^2 \ = \  - (\pa \h^1 / \pa x^1) + (\pa \h^2 / \pa x^3) -
(\pa \h^3 / \pa x^4) & \cr
{\dot x}^3 \ = \  (\pa \h^1 / \pa x^4) - (\pa \h^2 / \pa x^2) -
(\pa \h^3 / \pa x^1) & \cr
{\dot x}^4 \ = \  - (\pa \h^1 / \pa x^3) -
(\pa \h^2 / \pa x^1) + (\pa \h^3 / \pa x^2) \ . & \cr }
$$

\subsection{General structures}

Let us now consider a general hypersymplectic structure $\Ob$; we
denote a symplectic structure by $\om$ and a point in $M = \R^4$
by $x$; let $\om_x$ be the evaluation of $\om$ in $x \in M$. The
two lemmas below show that $\Ob$ is always equivalent to one of
the two structures considered above.

\medskip\noindent
{\bf Lemma 6.} {\it If $\om_x$ is unimodular, it belongs either to
$\S_+$ or to $\S_-$ .}

\medskip\noindent
\Proof Let $Y$ be the complex structure associted to $\om$.
We work in coordinates, and write in full generality $ Y (x )
=  ({\bf a \cdot K}) + ({\bf b \cdot H})$. Requiring $\om$ to be
unimodular, i.e. $Y^T Y = I$, we obtain two conditions: the
vanishing of off-diagonal terms reads $a_\a b_\b = 0$ $\forall
\a,\b=1,2,3$, so that $|{\bf a}| \cdot |{\bf b}| = 0$; setting
diagonal terms equal to one (together with the previous
condition) yields moreover $|{\bf a}|^2 + |{\bf b}|^2 = 1$. $\
\triangle$
\medskip

\medskip\noindent
{\bf Lemma 7.} {\it A hypersymplectic structure in $M = \R^4$ is
made either of positive type symplectic structures, or of negative
type symplectic structures, but in no case by symplectic
structures of the two kinds.}

\medskip\noindent
\Proof A hypersymplectic structure corresponds to a $su(2)$
algebra in the way discussed above. As also mentioned above, the
spans of the $\Kb$ and of the $\Hb$ correspond to $su(2)_\pm$
algebras, but no $su(2)$ algebra is generated by a mixture of
matrices belonging to $su(2)_+$ and $su(2)_-$ algebras. $\
\triangle$
\medskip

\subsection{Extension to 4n dimensions}

We stress that the standard structures are immediately extended
to structures in higher dimension. In the case $M = \R^{4n}$
(with euclidean metric), take {\it  block reducible} structures.
By this we mean that $\om_\a = (1/2) (J_\a)_{im} \d x^i \w \d
x^m$, and the $Y_\a = g^{-1} J_\a$ generate a representation of the
$su(2)$ algebra in $\R^{4n}$, which is the direct sum of
irreducible representations on four-dimensional subspaces.

In this case the matrices acting on each four dimensional block
will be either $\Kb$ or $\Hb$. We thus have, in block notation, $
J_\a = L^{s_1}_\a \oplus ... \oplus L^{s_n}_\a $, where $s_k =
\pm$, and $L^{(+)}_\a = K_\a$, $L^{(-)}_\a = H_\a$ (notice that
we could get equivalent hypersymplectic structures by orthogonal
changes of variables
on each $\R^4$ block). The analysis conducted in $\R^4$ does
apply on each block.

\section{Quaternionic oscillators.}

There is no need to stress the relevance and ubiquitous role of
(harmonic and nonlinear) oscillators in standard hamiltonian
mechanics; we want to discuss here the hyperhamiltonian oscillators,
with a view at the problem of integrable hyperhamiltonian systems
(we assume the reader is familiar with
hamiltonian integrable systems).

Our intuitive understanding of hyperhamiltonian integrable systems
will be that of systems which can be mapped to a system of
hyperhamiltonian oscillators.

We will consider systems in $M = \R^{4n}$ (with standard
euclidean metric), and standard (say positive type)
hypersymplectic structure; as the hyperkahler structure
induces in this case a quaternionic structure, we will speak
of {\it quaternionic oscillators}.

We will consider nontrivial systems with compact invariant manifolds,
and start by discussing the case $n=1$.

\subsection{The standard four dimensional case.}

The simplest nontrivial case of hyperhamiltonian dynamics is the
one where we have quadratic hamiltonians $\h^\a$, i.e. $\h^\a (x)
= (1/2) c_\a |x|^2$, with $c_\a$ real constants; in this case we
get ${\dot x}^i = c_\a K^\a_{ij} x^j$, which is easily integrated
(see the more general discussion below).

Let us actually write $\rho \equiv (1/2) |x|^2$, and consider the
class of nonlinear systems where $\h^\a (x) = \h^\a (\rho)$, i.e.
assume the $\h^\a$ are arbitrary smooth functions of $\rho$. We call these {\it quaternionic oscillators}.

In this case, write $A^\a = \d \h^\a / \d \rho$; we have $\grad
\h^\a = A^\a (\rho) x$, and the equations of motion (1) read
simply (see section 6)
$$ {\dot x}^i \ = \ \sum_{\a=1}^3 \, A_\a (\rho) \, (K_\a)^i_{~j}
\, x^j \ . \en{9} $$

Notice that $d \rho / dt  = \sum_{\a=1}^3  A^\a (\rho) \, [ x^i
(K_\a)_{ij} x^j ] =  0$; the last equality follows from $K_\a = -
K_\a^T$. Therefore $\rho$ (and hence $| x(t) |$) is a constant of
motion under any hyperhamiltonian flow for hamiltonians which
are  functions of $\rho$ alone.

As $\rho(t)=\rho_0$, we can on any trajectory rewrite (9) as
$$ {\dot x}^i \ = \ \sum_\a
\, c_\a^0 \, (K_\a)^i_{~j} x^j \ = \ \nu_0 (K_\a)^i_{~j} x^j \ , \en{10} $$
where $c_\a^0 = A_\a (\rho_0)$, and
$$ \nu_0 := \sqrt{(c_1^0)^2 + (c_2^0)^2 + (c_3^0)^2 } \quad ,
\quad K^i_{~j} \ = \ {1 \over \nu_0} \ \sum_{\a=1}^3  \, c_\a^0
\, (K_\a)^i_{~j} \ .$$

The solution to (10) is obviously $x(t) = \exp [K \nu_0 t] x(0)$;
expanding this in a power series in $t$ and using $K^2 = - I$, we
obtain immediately
$$ x(t) \ = \ \[ \, \cos (\nu_0 t) \, I \, + \, \sin (\nu_0 t)
\, K \, \] \ x(0) \ . \en{11} $$
This represents a uniform motion on a great circle -- identified
by the vectors $\xb_0 = x(0)$ and $\xb_1 = K x(0) $ -- of the
sphere $S^3$ of radius $r_0 = | x(0) |$. The frequency $\nu_0$ of
such motions will be the same for motions on the same sphere: it
depends only on the radius $r_0$ (note the $c_\a^0$ also depend
on $r_0$).

Therefore any sphere $S^3$ of radius $r_0 \not= 0$ is covered by
periodic circular motions, unless $\nu_0 (r_0) = 0$, all of them
with the same period $T_0 = 2 \pi / \nu_0$; in this way the
hyperhamiltonian flow (9) partitions $S^3$ into $S^1$ equivalence
classes (the dynamical orbits) and thus realizes a Hopf fibration
$S^3 / S^1 = S^2$ of the three-sphere \cite{Bre2}.

\subsection{The (4n)-dimensional case.}

Let us pass to consider $M = \R^{4n}$, again with standard
euclidean metric; we will use cartesian coordinates $\{ x^1 , ...
, x^{4n} \}$. We also define block variables $\{ \xi_1 , ... ,
\xi_n \} $ with $\xi_p \in \R^4$ corresponding to $x$ coordinates
in the $p$-th block, $\xi_p^i = x^{4(p-1)+i}$ (where $i=1,...,4$
and $p=1,...,n$); we also write $\rho_p = (1/2) |\xi_p |^2$.

We will consider the case where $\R^{4n}$ is equipped with a
standard block reducible hypersymplectic structure (see section
6), $J_\a = L_\a^{s_1} \oplus ... \oplus L_\a^{s_n}$. We have
$L_\a^+ = K_\a$, $L_\a^- = H_\a$.

We will now assume the hamiltonians depend only on the $\rho_p$
(we say we have a quaternionic $n$-oscillator):
$$ \h^\a (x) \ = \ \h^\a (\rho_1 , ... , \rho_n ) \ ; $$
we write the jacobian of the $\h$ with respect to the $\rho$
variables as $A^\a_p := \pa \h^\a / \pa \rho_p$.
In this case the equations of motion are (sum on repeated greek
and latin indices will be implied, except for the block index $p$)
$$ \xd^i \ = \ (J_\a)^{ik} \, \pa_k \h^\a $$
and can be written as (no sum on $p$)
$$ {\dot \xi}_p^i \ = \ A^\a_p (\rho_1 , ... , \rho_n ) \,
(L^{\s_p}_\a )^i_{~k} \, \xi^k_p \ . $$
Again the $\rho_p$ are constants of motion (no sum on $p$):
$$ {\d \rho_p \over \d t} \ = \ {\d \xi_p^i \over \d
t} {\pa \rho_p \over \pa \xi_p^i} \ = \ A^\a_p (\rho_1 , ... ,
\rho_n ) \ [ \xi_p^i \, (L^{\s_p}_\a)^i_{~k} \, \xi_p^k ] \ = \ 0 \ , $$
where the last equality follows from the antisymmetry of the
$L^{\s_p}_\a$.

Hence the matrices $A^\a_p$ are constant under the flow. If we
are given an initial datum $x(0)$, and thus the value of the
constants of motion $(\rho_1 =b_1 , ... , \rho_n = b_n )$, we can
write
$$ {\dot \xi}_{(p)} \ = \
\sum_{\a = 1}^3 \, c^\a_{(p)} \, (L^{\s_p}_\a ) \, \xi_{(p)} \ = \
\nu_{(p)} \, L_{(p)} \xi_{(p)}  \ , \en{16} $$ where $c^\a_{(p)}
= A^\a_p  (b_1 , ... , b_n )$, and
$$ \nu_p = \sqrt{ (c^1_{(p)})^2 +
(c^2_{(p)})^2 + (c^3_{(p)})^2 } \ , \ L_{(p)} = (1 / \nu_p )
\sum_\a  c_{(p)}^\a  L^{\s_p}_\a \ . $$

That is, on each block we have the same situation discussed in
subsection 1; notice that the frequencies $\nu_p$ depend not only
on the value $b_p$ of $\rho_p$, but also on the values $b_q$ of
the other variables $\rho_q$ ($p \not= q$).

\section{Discussion: the relation between hyperhamiltonian
and standard hamiltonian integrability.}

We would like to discuss the relation between hyperhamiltonian
integrability and standard hamiltonian integrability for the
class of systems considered here.

\subsection{Dimension four.}

Let us first of all focus on the case given by $\h^1 = |x|^2/2$,
$\h^2=\h^3=0$; this corresponds to two uncoupled and identical
harmonic oscillators with conserved energies $E_a = (1/2)
[(x^1)^2 + (x^2)^2]$ and $E_b = (1/2) [(x^3)^2 + (x^4)^2]$.

The solutions of nonzero energy $E = E_a + E_b = r_0^2/2$ describe
a circle $S^1$ lying on the sphere $S^3$ of radius $r_0$. When
$E_a$ and $E_b$ are both nonzero (i.e. both oscillators are
actually excited) these also lie on a torus $\toro^2 \ss S^3$, and
the circle $S^1$ corresponding to the solution is a combination of
the two fundamental cycles of the torus.

The cases $E_a = 0 , E_b \not= 0$ and $E_a \not= 0 , E_b = 0$
correspond to degenerate situations in which the common level set
of $E_a$ and $E_b$ is not a torus $\toro^2$, but is reduced to a
circle $\toro^1 = S^1$, which is just the trajectory of the
solution.

It should be recalled that the Hopf $S^3$ fibration can indeed be
described as a singular  fibration of $S^3$ in $\toro^2$ tori,
with two singular fibers, which correspond to the special cases
in which all the energy is on one oscillator and the other is not
excited; thus these two ways (hyperhamiltonian and standard
hamiltonian) of describing the situation are immediately related,
as it should be.
\medskip

Let us now come back to the general (nonlinear) integrable case
described by (9), whose solutions are given by (11); on each $S^3$
sphere of radius $r_0 \not= 0$, i.e. on each nonzero level
manifold\footnote{In this case we can speak of energy level
manifolds as the three hamiltonians depend on a single scalar
function $\rho$.} for the energy $E = \rho$ we can indeed reduce
to a two-oscillators description; see (9) and (10) above. Such a
system is integrable in the Arnold-\-Liouville sense, since the
set on which the fibration in tori is singular is of zero measure
in the phase space.

However, it should be noticed that in considering this system as
an integrable two-oscillator system, we are completely overlooking
the quaternionic structure of the system and of the whole class to
which it belongs. Also, this system is strongly degenerate if seen
in terms of two oscillators: indeed the two oscillators are in 1:1
resonance for all values of $H$, i.e. all values of the action
variables $I_1 = E_a$ and $I_2 = E_b$. Such a degeneration is of
course enforced by the quaternionic structure, and thus generic in
the frame of ``quaternionic oscillators''.

On the other hand, if we recognize the quaternionic structure and
the fact that we need therefore only the global constant of
motion $\rho$ to guarantee integrability (see the above
discussion, and the remarks below in this section), we have at
once a much stronger information on the structure of the system
and also need an easier construction to guarantee integrability.

The situation is similar to the one met when we represent a
quaternion by a pair of complex numbers (or a complex number by a
pair of real ones): this is possible and correct, but in this way
we are overlooking an additional and relevant structure, which we
must then introduce by suitable relations between complex (or
real) quantities.

Thus, in order to guarantee integrability in the sense of
standard hamiltonian mechanics we need two constants of motion
and we have to construct a system of two action and two angle
coordinates; using the quaternionic structure we only need one
constant of motion, i.e. $\rho$, and we have to construct a
system of coordinates in which to the ``action'' coordinate
$\rho$ are associated three ``angle-like'' coordinates. By
``angle-like'' we mean coordinates on the sphere $S^3$ which can
be seen as a generalization, from $S^1 \simeq {\bf C^1}$ to $S^3$
of the angular coordinates of standard hamiltonian mechanics; as
$S^3 \simeq {\bf H^1} \simeq SU(2)$ (here ${\bf H}$ is the
quaternion field and ${\bf H^1}$ the set of quaternions of unit
norm), these are of quaternionic nature. We call them {\bf spin
coordinates}.

Notice that the evolutions along spin coordinates do not commute;
thus the equivalent of the familiar integrable hamiltonian
evolution equations ${\dot I}_k = 0 $, ${\dot \phi_k} = \om_k
(I)$, related to the abelian group ${\bf T}^2$, is now given by
(9), (10) or, more intrinsically, by ${\dot I} = 0$ ($I \equiv
\rho$), ${\dot \psi} = \a (I)$, where $\psi$ represents
coordinates on the group $SU(2) \simeq S^3$, and $\a (I) \in
su(2)$ is an element of the algebra $su(2)$, constant on each
level set of $I \equiv \rho$. This more involved (and not
separable) structure is unavoidable, due to the non-abelian
nature of $SU(2)$.

\subsection{Higher dimension.}

In the standard hamiltonian integrable case with $m$ degrees of
freedom, i.e. for $m$ hamiltonian oscillators (say all of them
excited) we have invariant $\toro^m$ tori, and the solutions will
cover densely $\toro^k \ss \toro^m$ tori, with $k \le m$
depending on the rational relations between the frequencies of
different degrees of freedom on the given $\toro^m$; in the
hyperhamiltonian integrable case (for $n$ quaternionic
oscillators) we have a similar situation, as we now discuss.

First of all we remark that, since $\rho = (\rho_1 , ... , \rho_n
)$ are constants of motion, the common level sets of the $\rho_p$
are invariant manifolds under the dynamics we are considering;
these level sets $\rho^{-1} (b_1 , ... , b_n) $ will be, when all
the $b_p$ are nonzero, manifolds $$ S^3 \x ... \x S^3 \ = \
\vacca^n \ ; $$ notice that these $\vacca^n$ represent a
generalization of tori: in the same way as $\toro^n$ is the
topological product of $n$ (distinct)  $S^1$ factors, $\vacca^n$
is the topological product of $n$ (distinct)  $S^3$ factors. If
$k$ out of the $n$ numbers $b_p$ are zero, the level set
$\rho^{-1} (b_1 , ... , b_n) $ will be a $\vacca^{n-k}$ manifold.

We will denote the trajectory with initial datum $x(0)$ as $\ga
\ss \R^{4n}$. The previous discussion shows that the projection
of $\ga$ to each $\R^4$ block, given by $\xi_{(p)} (t)$, will be
periodic.

If all the frequencies $\{ \nu_1 , ... , \nu_n \}$ are rational
with respect to each other, the full solution in $\R^{4n}$ will
also be periodic, i.e. $\ga \approx S^1$; if $m$ degrees of
freedom are excited (i.e. there are $m$ nonzero $b_p$'s), this
$\ga$ will also be a submanifold of the invariant manifold
$\vacca^m = \rho^{-1} (b_1 , ... , b_n)$.

If $m$ degrees of freedom are excited  and the frequencies
corresponding to $b_p \not= 0$ split in $k \le m$ sets, each
$\nu_p$ being rational with respect to frequencies in the same
set and irrational with respect to frequencies in different sets,
the solutions $\ga$ will densely cover $\toro^k$ tori.

 These $\toro^k$ will be submanifolds of $\vacca^m$, and we can always
choose the generators $S^1$ of $\toro^k$ so that each generator
lies in a different generator $S^3$ of $\vacca^m$. Indeed each
generator of $\toro^k$ will be a linear combinations of the
projections of $\ga$ to the blocks corresponding to each rational
subset of frequencies; we can choose it to be just a single
projection and thus to be in a $S^3$ factor of $\vacca^m$.

\section{Final remarks}

In this final section we briefly present some additional remarks to put our work into perspective and mention directions of future developement. We thank an unknown referee for raising the problem discussed in point 4 below.
\medskip  

{\bf (1)} First of all, it should be stressed that here we were mostly interested in the local structure of this hyperhamiltonian dynamics, and we have not considered problems arising from the global structure of the hyperkahler manifold $M$ on which it is defined. Locally, any such $M$ is isomorphic to $\R^{4n}$, so that we could have limited to consider these spaces (as in sections 6--8). 

However, as in the standard hamiltonian case, most of our construction will extend to more general hyperhamiltonian manifolds, so that in our general discussion (sections 1--5) we preferred to deal with a generic hyperkahler manifold, pointing out where our discussion requires to work chart by chart.

Focusing on local properties means, of course, that we are not concerned with the geometrically most interesting recent results on hyperkahler manifolds and their global structure (which is also relevant in connection with Physics); see the references given in the introduction, and in particular \ref{Qconf}, for an overview of these. 

{\it A fortiori} we are not providing any new insight into hyperkahler geometry nor are we providing any new nontrivial hyperkahler manifold.
We actually needed only the very basic definitions of hyperkahler geometry; we supposed that a hyperkahler manifold $M$ is given, and we defined a dynamics on $M$ related to the choice of three hamiltonian functions. 
\medskip

{\bf (2)} We should also notice that no attempts to generalize hamiltonian dynamics in the direction proposed here seems to be present in the rapidly growing literature on hyperkahler manifolds (mostly devoted to their geometry and construction of nontrivial examples). 
A somehow orthogonal approach to a hyperkahler generalization of the structure of standard hamiltonian mechanics, focusing on Poisson structures, was suggested by Xu \cite{Xu}. 
\medskip

{\bf (3)} We also mention that in field theory considerations of multiple symplectic structures is suggested by covariance requirements and lies at the basis of the de Donder - Weyl formalism, as discussed in detail in \cite{GIM}, who call this ``multisymplectic field theory''; however our approach (limited to mechanics) seems -- at least at the present stage -- not related to this theory.
\medskip

{\bf (4)} As mentioned in the introduction, a most important result in hyperkahler geometry concerns the construction of nontrivial hyperkahler manifolds via a moment map procedure starting from a (possibly trivial) hyperkahler manifold equipped with a Lie group action \cite{Bie,Hit2}. It is natural to ask what happens when a (covariant) hyperhamiltonian dynamics is defined on the first manifold, i.e. how the dynamics descends to the quotient.

Let $(M,g;\om_\a)$ be a hyperkahler manifold of dimension $m$. Assume that there is a compact Lie group $G$ (we denote by $\G$ the Lie algebra of $G$ and by $\G^*$ its dual) acting freely on $M$ and preserving its metric $g$ and the three forms $\om_\a$ (thus acting triholomorphically); this defines three moment maps $\mu_\a : M \to \G^*$, or a map $\mu : M \to \G^* \otimes \R^3$. It is known \cite{Hit2} that the quotient metric on $N = \mu^{-1} (0) / G$ is hyperkahler. We denote by $\b_\a$ the reduction of $\om_\a$ on $N$.

Let $\h^\a$ be three $G$-invariant smooth functions $\h^\a : M \to \R$, $\h^\a (x) = \h^\a (g x)$ for all $g \in G$ and $x \in M$, and $X$ be the hyperhamiltonian vector field on $M$ corresponding to these; we recall that $X$ is given by $ X = \sum_\a X_\a$ with $X_\a$ identified by $X_\a \interno \om_\a = \d \h^\a$ (no sum on $\a$). 

However, each $X_\a$ is a hamiltonian vector field, generated by the hamiltonian $\h^\a$, with respect to the symplectic structure $\om_\a$. Thus, each $X_\a$ descends to a hamiltonian vector field $W_\a$ on $N$, by standard symplectic reduction. In other words, for each $\a$ there is a smooth function $\K^\a : N \to \R$ such that $W_\a \interno \b_\a = \d \K^\a $ (no sum on $\a$). 
This shows at once that $X$ descends to a hyperhamiltonian vector field $W = \sum_\a W_\a$ on the hyperkahler quotient $N$.
\medskip

{\bf (5)} Physically, one should consider generalizations of the present approach in at least two directions: on the one hand, one would like to consider pseudoriemannian rather than riemannian manifolds; and on the other hand one should consider quantum version of the theory. It appears that both of these are feasible, and we will report on these matters in a separate note.
\medskip

{\bf (6)} Finally, we would like to point out that the dynamics introduced here can be obtained in a completely different and quite interesting way. One can look at standard Hamilton equations in terms of complex analysis, and extend them from the complex to the quaternionic case; one obtains then exactly the equations introduced here, as discussed in \cite{MT}.

\vfill\eject

\section*{Appendix. Ideals of differential forms.}

In our discussion of the variational formulation of the hyperhamiltonian equations of motion, we used some concepts from the theory of ideals of differential forms (here called just ideals, for short). This is maybe less widely known that the other tools used in the paper, so we collect here some definitions for convenience of the reader; see \cite{Car,God} or \cite{Arn2} for further detail.

\medskip\noindent
{\bf Definition A1.} {\it Let $M$ be a smooth $N$-dimensional manifold, and $\J_k \ss \Lambda^k (M)$, for $k=0,...,N$. The subset $\J = \bigcup_{k=0}^N \J_k \ss \Lambda (M)$ is said to be an {\rm ideal of differential forms} iff: 
{\bf (i)} $\eta \in \J$, $\psi \in \Lambda (M)$, $\Rightarrow \eta \w \psi \in \J$; 
{\bf (ii)} $\beta_1,\beta_2 \in \J_k$, $f^1, f^2 \in \Lambda^0 (M)$, $\Rightarrow f^1 \beta_1 + f^2 \beta_2 \in \J_k$.}

\medskip\noindent
{\bf Definition A2.} {\it Let $i : S \to M$ be a smooth submanifold of $M$; $S$ is said to be an {\rm integral manifold} of the ideal $\J$ iff $i^* (\eta ) = 0$ for all $\eta \in \J$.}
\medskip

The ideal $\J$ is said to be {\it generated} by the forms $\{ \eta^{(\a)} , \a= 1,...,r \}$ (with $\eta^{(\a)} \in \J$) if each $\phi \in \J$ can be written as $\phi = \sum_\a \rho_{(\a)} \w \eta^{(\a)}$ for a suitable choice of $\rho_{(\a)} \in \Lambda(M)$, 
$\a = 1,...,r$.
If $\J$ is generated by $\{ \eta^{(\a)} , \a= 1,...,r \}$, then $i : S \to M$ is an integral manifold for $\J$ iff $i^* (\eta^{(\a)} ) = 0$ for all $\a = 1,...,r$.

Given an ideal $\J$, we associate to any point $x \in M$ the subspace $D_x (\J) \ss \T_x M$ defined by $ D_x (\J) := \{ \xi \in \T_x M \ : \ \xi \interno \J_x \ss \J_x \} $.

\medskip\noindent
{\bf Definition A3.} {\it If $D_x (\J)$ has constant dimension, the ideal $\J$ is said to be {\rm non singular}, and the distribution $D (\J) = \{ D_x (\J) , x \in M\}$ is its {\rm characteristic distribution}; any vector field $X \in D (\J)$ is said to be a 
{\rm characteristic field} for $\J$.}
\medskip

The following proposition is easy to prove (e.g. using the local coordinates introduced in section 44 of \cite{Car}, or directly from definitions above) and is used in remark 8. 

\medskip\noindent
{\bf Proposition A1.} {\it Let $\J$ be generated by forms of degree $k$, and let $i : S \to M$ be an $r$-dimensional integral manifold of $\J$, with $r < k$. Let $X$ be a characteristic vector field for $\J$, and let $X$ be nowhere tangent to $i (S)$. Let $\phi_t $ be the local one-parameter group of diffeomorphisms generated by $X$. Then the  $(r+1)$-dimensional manifold $(- \varepsilon , \varepsilon ) \times S \ni (t,x) \mapsto \phi_t (x) \in M$ is an integral manifold of $\J$.}

\vfill\eject

\end{document}